\documentclass[conference]{IEEEtran}

\ifCLASSINFOpdf
  \usepackage[pdftex]{graphicx}
\else
\usepackage[dvips]{graphicx}
\usepackage{multirow}
\fi

\begin{document}
%
\title{SecSip: A Stateful Firewall for SIP-based Networks}

\author{\IEEEauthorblockN{Abdelkader Lahmadi and Olivier Festor}
\IEEEauthorblockA{INRIA Nancy - Grand Est Research Center, Villers-L\`es-Nancy, France\\
Email: \{Abdelkader.Lahmadi,Olivier.Festor\}@loria.fr}
}


%


\maketitle

\begin{abstract}

 SIP-based networks are   becoming  the de-facto  standard  for voice,
 video and instant messaging services.  Being  exposed to many threats
 while   playing  an major role     in the operation  of essential
 services, the  need for dedicated   security management approaches is
 rapidly increasing. In this paper we  present an original security
 management approach based    on a specific vulnerability  aware SIP 
 stateful firewall.  Through known  attack descriptions, we illustrate
 the  power of the configuration  language of the firewall which uses
 the capability to    specify stateful objects  that  track  data from
 multiple SIP elements within  their lifetime. We demonstrate  through
 measurements on a real implementation  of the firewall its  efficiency
 and performance.

\end{abstract}

\begin{IEEEkeywords}
SIP, VoIP, Security, Firewall
\end{IEEEkeywords}

%
\IEEEpeerreviewmaketitle

\section{Introduction}

 The Session Initiation  Protocol (SIP) \cite{rfc:sip} has established
 itself among the most important Internet protocols. It is designed to
 establish,    modify, and terminate    a  session of  application
 services.  SIP is  currently used  in many  popular services such  as
 Voice over IP (VoIP), Instant Message (IM), Presence Service and even
 File Transfer.  In the near future, it is expected that SIP will play
 an  essential  role  in   the   next-generation telephony   networks.
 Traditional telephony based on  PSTN networks was well secured, since
 it  is  based on  close  environments,  where calls  are  carried by
 dedicated lines and managed  by operator-owned devices.  To carry  calls,
 SIP-based   service architectures    use  the   Internet and   expose
 themselves to all kinds of attacks ranging from Distributed Denial of
 Service to toll-fraud,  vishing or eavesdropping \cite{voipsa}.

 Offering an  efficient  security     management framework  for    SIP
 infrastructures is  becoming a challenge to the  success and the wide
 deployment  of VoIP services. We  believe that one essential building
 block of  such  a security management   framework is a  dedicated SIP
 firewall. It  must be dedicated because the  use  of generic IP-based
 firewalls are inefficient to  address and mitigate most attacks against
 SIP services. IP-level firewalls actually have two major drawbacks :

\begin{itemize}

 \item  they do not   address  the SIP  protocol  semantics.  The  SIP
 protocol messages are text based, and the various fields of a message
 are exploited  to  carry out different kinds of   attacks. For
 example, an attacker  can easily employ the  SIP BYE  request to tear
 down a session, without any violation of an IP level firewall rule.

 \item   they  do    not  allow to   consider    per  device  specific
 vulnerabilities.        Different    devices   have       different
 vulnerabilities. Being   able to  protect them   in  an efficient way
 requires both  the  knowledge    of   the  devices and    a   precise
 specification of the specific  vulnerabilities. This is not  provided
 by IP level firewalls.
\end{itemize}

 To remedy these shortcomings, we have designed and implemented a SIP defense system that
 does support  an in-depth  message analysis together with a SIP protocol
 state   tracking function.   It  is  also  required that  the
 designed SIP defense  system  satisfies the following properties:  it
 must be fast, accurate, induce low overhead and be safe. Our approach
 to this  problem  is an application  level  firewall implemented as a
 ``bump in  the wire'' device:  it  intercepts SIP messages, evaluates
 their  safety before forwarding them  to their  destinations.  If the
 message or the transaction to which a message belongs is unsafe, our system
  SecSip blocks it, thus protecting the device from the threat.

\begin{figure*}[htp]
 \begin{center}
 \scalebox{0.3}{\includegraphics{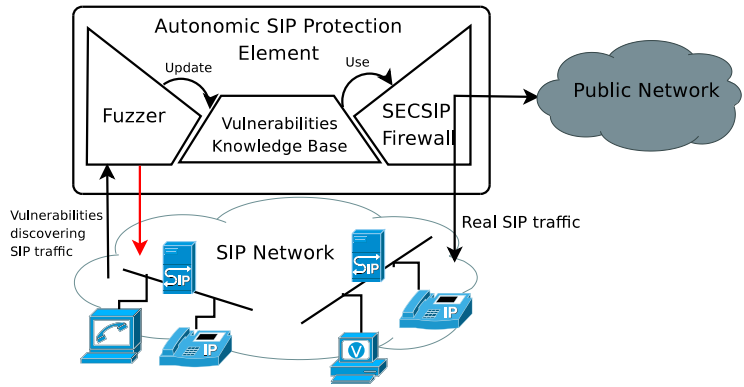}}
 \caption{Overall approach combining the SecSip firewall and fuzzing tools.}
 \end{center}
\label{fig:autonomic} 
\end{figure*}

 Our approach to protect SIP networks from diverse vulnerabilities
 is depicted in Figure \ref{fig:autonomic}. The fuzzing module allows 
 the discovering of per device SIP vulnerabilities in the network. It then updates 
 the vulnerability knowledge base that shares with the firewall named SecSip. 
 In a second stage, SecSip translates this vulnerability into defense rules
 to protect devices from the threat.  
   	
 The SecSip environment   combines  two key techniques.   First,  it   uses a
 rule-based  engine to execute   rules that model SIP vulnerabilities.
 These rules are executed against  SIP protocol messages, transactions
 and dialogs.  Second, it monitors the SIP protocol to enable stateful
 semantic tracking through stateful objects claimed by vulnerabilities
 defense rules. Hence, rules that  model SIP vulnerabilities are based
 on both protocol behaviour and attack signatures.
 
 The system can operate  either as a service  deployed in  the network
 infrastructure or as a  client-side  protection tool. While each  has
 its merits  as we will explain  later, we focus in  this paper on the
 network service deployment model on which performance is a key issue.

 The   remainder  of  the  paper  is  organised  as  follows.  Section
 \ref{sec:related} gives an overview on SIP vulnerabilities as well as
 on achievements    in  the   area of     firewalls  and  the  associated
 specification languages. Then, we  discuss the design  space of a SIP
 defense system in section  \ref{sec:design}.  The SecSip runtime  and
 language are described in section \ref{sec:architecture}. Their evaluation 
is performed in   section  \ref{sec:evaluation}.   Finally,  we
 conclude on the contribution and outline some  future    work  in     section
 \ref{sec:conclusion}.

\section{Related work}
\label{sec:related}

\subsection{SIP exploits and vulnerabilities}

 VoIP  networks are subject to  many types of attacks.   A rich set of
 existing work   \cite{zhang:07,state:07,Chen:06,Sip:threats}     has
 addressed SIP vulnerabilities and exploits to examine how they can be
 efficiently used to compromise the reliability and trustworthiness of
 SIP-based VoIP systems.  In \cite{zhang:07}, authors focus on billing
 attacks The SIP protocol is  also exposed to traditional DoS  attacks
 \cite{Habib:03} like network  bandwidth  and  OS/firmware attacks  to
 exhaust  available resources.  Furthermore,   SIP comes with  its own
 specific   DoS    attacks.    These   attacks   are  illustrated   in
 \cite{Chen:06} where the author presents  stateful solutions based on
 finite-state machines  for SIP  transactions to  detect  them.  These
 two contributions, have mainly illustrated SIP exploits rather than the
 specification  of the  vulnerability that allows   such attacks.   In
 \cite{Sip:threats},  authors enumerate  SIP attacks and  identify the
 vulnerabilities that  causes them. The lack  of authentication is for
 example, the major  cause of signalling attacks  like BYE, CANCEL and
 Re-INVITE.

\subsection{Application Level Firewalls and their language}

 There is a   lot of literature on   application level firewalls   and
 intrusion  detection systems  \cite{modsecurity,snort,bro} devoted to
 common protocols   like HTTP, SMTP,  etc.  The  SecSip language is
 inspired by  the ModSecurity \cite{modsecurity} approach which allows
 HTTP   traffic  monitoring and  filtering,   with real-time intrusion
 detection.   Snort   and  Hogwash \cite{snort}  are   other intrusion
 detection systems that  target   mainly  the network    level.   They
 recently started to support some SIP exploits such as INVITE flooding
 attacks in a very limited way.

 VoIP firewalls and more  specifically those addressing the defense of
 the SIP protocol, are still in early stages and there is only limited
 work    published     \cite{fielder:07,Wu:04,sengar:06,Chen:06}. 
 The authors of \cite{Wu:04} propose a solution for  stateful intrusion
 detection called \textit{SCIDIVE}.    The system relies on a
 stateful engine    that determines the  current state   from multiple
 packets involved  in   the  same  session.   The  system  also   uses
 cross-protocol  detection to    verify the   consistency between  two
 protocols involved in  the same VoIP session, mainly  SIP and RTP. 
The goal   of their work are  similar  to SecSip and  shares
 the stateful feature.
 
 VoIP defender \cite{fielder:07}  is a SIP-based security architecture
 designed to monitor, detect, analyse and  counter attack. 
 The nature of the employed  detection
 scheme   (stateful or stateless) is not clearly defined in the publication. In addition, no details are  provided in 
 about the language used to build defense rules and how it can
 be used for SIP.

 Our    work is complementary  to    existing SIP vulnerability  and
 exploit discovery   tools.  SecSIP uses  the output from vulnerability
 discovery systems  like KIF \cite{kiph} to  close the  defense loop by  enforcing
 security policies  against attacks  exploiting these vulnerabilities.
 We also take profit from  existing intensive literature dedicated  to
 application level firewalls, to instantiate the system.

\section{SIP defense design space}
\label{sec:design}

 The SIP protocol is transaction-based. Each transaction consists of a
 request that invokes a particular method, or  function, on the server
 and  at least  one  response  \cite{rfc:sip}.  Attackers usually  use
 malformed  SIP messages within a    transaction to compromise a   SIP
 entity.  They  also  employ legitimate   messages   to attack a   SIP
 infrastructure (e.g. redirect calls, end a session, cancel invitation
 and update session parameters).  Therefore, a SIP defense system must
 be able  to defend a SIP  infrastructure  against both malicious (but
 legitimate) transactions and malformed  messages. 

 The underlying  approaches  we do consider   fall into two  categories: proactive and
 reactive.    Proactive  defense prevents  malicious transactions  and
 malformed packets   from  reaching  the intended  victim.   A  common
 proactive  approach to   identify malicious behaviour  is to  record
 interacting SIP   states, objects and  messages.   In this  case, the
 approach is stateful.  A proactive approach may operate anywhere
 in  the network perimeter.  However,  if located at
 the victim  side such an approach  becomes  useless, since a   denial of service attack
 damage still occurrs while   the defense system  tries to prevent
 it. A reactive defense approach generates    an inoculation   in  response to     an
 attack. This response  will protect SIP  devices.  An example of such
 inoculation is to deploy patches to  eliminate a bug exploited by the
 attacker.   There are many  additional  examples of reactive  defense
 approaches including   intrusion  prevention    systems,  statistical
 analysis,  attack signatures, reactive  address   blacklisting, etc.  These
 solutions   attempt     to recognise   post-attacks    and    take  a
 counter-measure later.

 The effectiveness of  each approach depends  on the type of attack. A
 proactive approach is suitable to  cover compromise attacks like toll
 fraud, unwanted calls  and messages, \ldots \cite{Sip:threats}.  This
 type of  attack needs  very  few SIP messages  to cause  damage.  The
 proactive  approach needs   to  identify  malicious SIP objects   and
 prevent them from reaching the victim.  The  defense strategy can be
 exercised  either   in the  network  perimeter   or at  the  victims'
 location. Denial of service attacks keep the victim unaffected if applied
 at the network level rather than on end-systems.

\begin{table*}[ht]
\caption{Design space of SIP defense solutions.}
\label{tab:space}
\centering
\begin{tabular}{|l|l|l|l|l|}
\hline
Attack type &  \multicolumn{3}{|c|}{Defense approach} & Criteria \\ 
\hline 
& Proactive at SIP device & Proactive in SIP network & Reactive in SIP network & \\
\hline 
\multirow{3}{*}{Compromise attacks} & High & High & Not useful & Assumptions \\
 & High & High & & Effectiveness \\
 & Low  & High & & Complexity \\
\hline
\multirow{3}{*}{Deny of service attacks} & Not useful & High & Medium & Assumptions \\
                                             &              & High & Medium & Effectiveness \\
                                             &              & High & Medium & Complexity \\
\hline 
Examples & Firewalls & Firewall, NAT & IDS &  \\
\hline
\end{tabular}
\end{table*}

 The  design  space of SIP  defense  solutions is  summarised in Table
 \ref{tab:space}.  We observe that  a proactive SIP defense in network
 is more efficient than   others, but also more challenging to build. It needs high
 assumptions about  safety  and   complexity.  However  it  is   more
 efficient to protect SIP-based networks.

\section{Architecture and components}
\label{sec:architecture}

\begin{figure}[ht]
\centering
\scalebox{0.7}{
\includegraphics{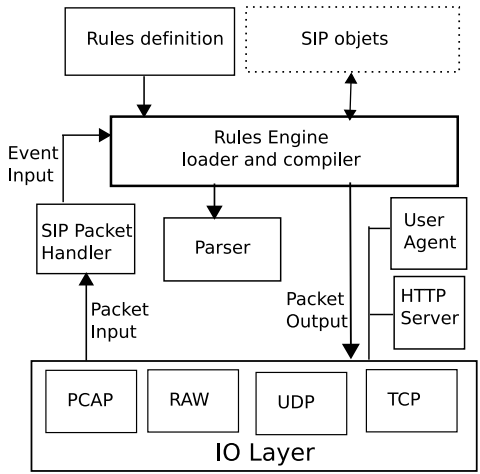}
}
\caption{Overview of SecSip architecture.}
\label{fig:arch}
\end{figure}

 Our approach to defend SIP-based  networks relies on the insertion of
 a  proactive point of defense between  a SIP-based network of devices
 (servers, proxies, user agents) and the open Internet. Therefore, all
 SIP traffic is inspected and analysed before it is forwarded to these
 devices.   Figure  \ref{fig:arch}   depicts  the   proposed    SecSip
 architecture that integrates four major components:
 Input/Output Layer, Stateful Inspection Layer,  SIP Packet Handler and
 Rule Compiler and Optimizer.

Each component is described in details in the following sections.
\paragraph{Input/Output Layer}

 The input and  output features provide  the service able  to capture,
 inject, send and  receive SIP packets from  and to the network.  They
 are configured according to the deployment  mode of the firewall.  In an
 in-line mode, the capture feature is active on the incoming interface
 (from the Internet). If the packet is safe, the firewall, will inject
 it,  using raw  sockets, on the  SIP  network while  keeping the same
 source and destination addresses.  Features to accept TCP connections
 and UDP traffic for further deployment modes is also supported.

\paragraph{SIP Packet Handler}

 Intercepted packets  are moved to the SIP  Packet  parser module. The
 main function of this module is to extract  different fields within a
 SIP message. Each field within a SIP message  is the composition of a
 key  that represents   the field name   as  defined  by the   SIP BNF
 \cite{rfc:sip} and its respective value. While parsing a sip message,
 the parser builds a data  tree that represents  the SIP packet.  Each
 node of the parse tree represents a SIP field  defined by a numerical
 identifier. The  parse  tree  node also contains  useful  information
 about  the field such as  type, length,  starting and ending offsets
 within the SIP  packet, \ldots This information is used later  by
 the stateful inspection layer to check the various SIP fields.

\paragraph{Rules Compiler and Optimizer}

 The   core of the SecSip  firewall  is its  rule   engine. It has the
 critical task to  process   defense rules  against  SIP messages  and
 transactions. When initializing, the  rule engine starts loading and
 parsing rules to identify  different targets, operations  and actions
 within each  rule.  Each rule is transformed  into a specification. A
 specification  is a data structure that  holds the names and values of each
 object of  the   rule. Then, rule   specifications are   compiled  to
 identify  the targeted fields from  the parse tree, the pre-registered
 operations and actions. 

 During rule parsing, the  engine creates  stateful objects to  store
 SIP  dialogs related data.  Each defined  rule set is  attached to a
 SIP  transaction state machine  type as  specified in \cite{rfc:sip}.
 The  rules engine is thus  able  to store  defense rules according to
 their transaction state machine types.  It uses a hash table to store
 rules where each entry is defined by the rule's transaction type.
 After being stored, an optimization component traverses the rules hash table 
 and starts scheduling their execution. This component specifies how rules
 will be executed when a SIP message arrives to the firewall. 
 Rules are ordered in a scheduling list according to their referenced objects.
 At the top of the list are the rules that declare objects and acts on their values.
 At the bottom of the list are rules that have many matches and have disruptive actions 
 on a SIP message. 
 In the example depicted in Figure \ref{fig:example_rules}, we have two SecSip rules. The first rule 
declares and updates a counter that counts the number of received INVITE messages.
The values 10 and 60 denote that we need to decrease the counter value by 10 each 60 seconds.
The second rule will drop SIP messages when the counter is greater than 80.
\begin{figure*}[htc]
\begin{center}
\begin{tabular}{ll}
R1: & secsip "FIELDS:sip.method" "\^{}INVITE" declare:rate=counter[10;60] \\
R2: & secsip rate "@eq 80" drop \\
\end{tabular}
\end{center}
\caption{A sample of SecSip rules}
\label{fig:example_rules}
\end{figure*}

 Even, if the rules are reversed in the configuration files and the administrator writes R2 before
 R1. The SecSip optimization component will schedule R1 followed by R2 in its scheduling list since 
 R2 references the variable rate that is declared and updated by R1.

\paragraph{Stateful Inspection Layer}

The inspection layer executes the appropriate rules on each received SIP packet
without any buffering, even if the SIP message is incomplete. 

\begin{figure}[htc]
\centering
\scalebox{0.7}{
\includegraphics{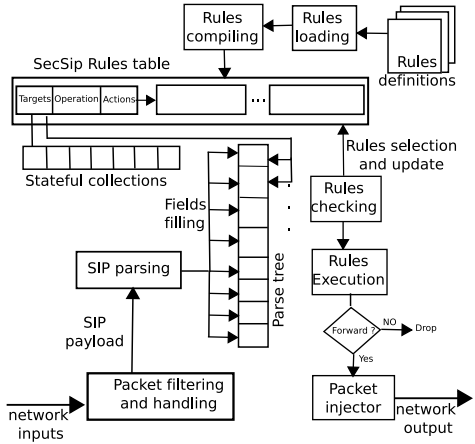}
}
\caption{SIP message traversal through SecSip runtime.}
\label{fig:flow}
\end{figure}

 Figure \ref{fig:flow} shows how a  SIP  message traverses the  SecSip
 runtime and gets analyzed.  First a SIP message  is captured from the
 network interface   by the  IO   layer and delivered  to  the parsing
 module. Then, the packet is parsed according the pre-registered fields
 of    the   SIP   parse   tree  that     follows   the  SIP    format
 specification.  These parsed fields    will fill the  parse tree  and
 synchronize    the stateful objects   defined  by  the SecSip defense
 rules.  
 SecSip stateful objects are stored to a certain type of container
 specified in a specific rule. The SecSip language provides three types of containers:
 set, list and bag. A set is an unordered collection of objects without repeated values. A list is an ordered collection of objects. 
A variation of a set is the bag. It allows repeated values and multiple objects. 
 The different containers referenced by stateful objects are stored within a hash table. 
Where each container has a key
that is a SIP dialog identifier defined by the triple (Call ID, To Tag, From Tag).
 Based on  the  current  state of the   SIP   session and the
 direction of the  packet, the corresponding  matching set of rules is
 invoked   to be  processed.  Rule  processing  will take  considered
 actions that may  lead to a decision of forwarding the packet to its
 destination or to drop it.

\subsection{The SecSip language}

 The main feature of the stateful firewall is its language designed to
 model SIP vulnerabilities.  We here  define a SIP vulnerability as  a
 flaw in one of its objects  where execution may  go wrong and violate
 the intended semantics of the SIP protocol  (i.e., the protocol state
 machine and message formats).

This leads to the following definitions  :

 \paragraph*{Definition 1}  $S$ is the  set   of all  possible  object
 states in a SIP interaction.

 \paragraph*{Definition 2} A  SIP vulnerability is a  tuple $v=(f,P)$,
 where $f$ is a state transition function that defines post-conditions,
 and $P\subseteq   S$  is a   set  of SIP  objects states   satisfying
 pre-conditions.

 A  vulnerability contains a set $P\subseteq  S$,  where P defines the
 required SIP object state for the vulnerability  to exist, also known
 as  pre-condition.   The   post-condition  function $f$   expresses a
 transition from one SIP object state to another.

 \paragraph*{Definition   3}  A SecSip   rule  is   modeled as a  tuple
 $i=(v,S^{'},A)$, where $v$ is a vulnerability, $S^{'}$ is a SIP state
 defined  by $S^{'}=f_v(P_v)$ and  $f_v$ is the vulnerability function
 associated to $v$.

 In this setting if $S^{'}$ is detected by the post-condition function
 $f_v$ , the SecSip runtime triggers the set of actions $A$, before 
enabling $f_v$ to occur on the real network.

 A  SIP object contains properties that  describe  network entities and
 logical relationships.   Logical relationship describes communication
 and trust patterns between SIP  network entities.  We identify  three
 types  of entities for which SIP objects are necessary :  messages  and
 their  fields,  transactions  and  dialogs.   In the  SecSip defense
 language, a user-defined SIP object follows the dot notation. It's syntax form   
 is as follows:  \textit{Object::= ObjectIdentifier [*('.' ObjectIdentifier)]}

 A user-define SIP object describes the  SIP protocol properties over
 its lifetime.   In  the  SecSip language,  we define
 several kinds of  SIP  objects with different  lifetimes  and scopes.
 Table \ref{tab:objects} summarizes the SIP objects defined by the SecSip
 language. These objects are employed   by the   user-defined rules  to detect   SIP
 vulnerabilities.

\begin{table}
\begin{center}
\begin{tabular}{|l|l|l|}
\hline
Object scope & Lifetime  & Type\\ 
\hline\hline
message-object & message & stateless\\ 
\hline
transaction-object & transaction & stateful \\
\hline
session-object  & session & stateful\\
\hline
\end{tabular}
\end{center}
\caption{The lifetime and scopes of SecSip language objects}
\label{tab:objects}
\end{table}

Stateful objects are defined using  the \textit{hold} instruction. A
stateful object has predefined properties when it is created.
 
 The objects related   to a  SIP  message  track  the values  of   its
 fields.   These objects are   stateless since they are re-initialized
 with each message. By default the SecSip runtime provides a set of stateless objects mapped to the parsed fields of a SIP message. These objects are defined by the \textit{FIELDS} or \textit{BODY} 
 identifiers mapped to the header or the body parts of a SIP message.
 The identifier is followed by the name of the object that follows a dot notation.
 For example, to access the \textit{From} specific  field of a SIP message,
 we use: $FIELDS:sip.from$.

 Transaction   related  objects are  used  to  record  data  across the
 lifetime of  a      transaction, spanning a     request  and multiple
 responses. Within the SecSip  runtime, a transaction is  identified by
 the  combination of the  \textit{BranchID} and the \textit{CSeq} command value.

 A dialog is defined by the \textit{Call-ID} together with the \textit{From IP} and \textit{To IP}. It spans multiple
 transactions.  Objects related to a dialog  are stateful and provide
 data across the lifetime of  the dialog.  For example, to initialize
 a  stateful object that tracks  the values  of the \textit{From} field in all messages within a dialog, we use the following:

\begin{small}
\begin{verbatim}
hold:FROM_LIST=set[MESSAGE_HEADERS:sip.from]
\end{verbatim}
\end{small}

In the above example, the statement defines a stateful object $FROM\_LIST$ that
holds the values of the stateless object \textit{sip.form} from the message-object \textit{FIELDS} over 
all messages within a dialog. We note, that $FROM\_LIST$ is a user defined object that will hold stateful data.
However, the \textit{FIELDS:sip.form} is a predefined object in the SecSip language.

To illustrate the ease of use of the SecSip language, we present two real attacks against SIP protocol
on which we will illustrate the defense specification.

\paragraph{DoS attacks detection}

\begin{figure}[htc]
\centering
\scalebox{0.4}[0.3]{
\includegraphics{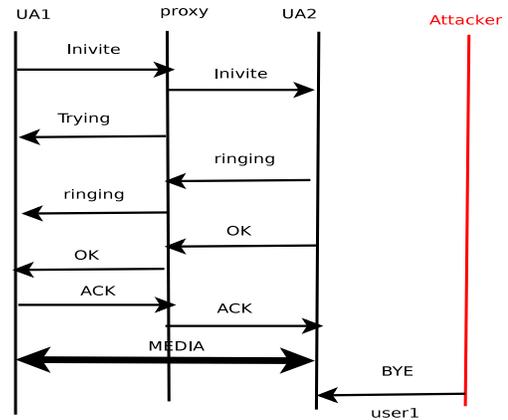}
}
\caption{Illustration of the BYE attack.}
\label{fig:bye-attack}
\end{figure}

 A  well known example  of DoS attack against  the SIP protocol is the
 BYE-attack  \cite{Sip:threats}.   In this scenario,   the  aim of the
 attacker is to teardown a VoIP session between two UAs.  To this end,
 it sends  a faked BYE message to  one UA on behalf  of  the other UA.
 When the  targeted UA receives the fake  BYE  message, it prematurely
 tears down the  established call  assuming  that is requested  by the
 partner     UA.      This    attack    is   illustrated    in  Figure
 \ref{fig:bye-attack}.

 The detection of this attack, needs  to track stateful objects within
 a  SIP  session\footnote{i.e.  a sequence    of multiple SIP messages
 exchanged between two  or more SIP entities.}.   In this scenario, we
 start  with recording the values of   the  \textit{From} field used by
 SIP  messages within   a  session  in the   user   defined stateful object 
 $from\_list$.  We  also  track the IP  addresses in  the messages and
 record them under the  main stateful object \textit{from\_list}  with a child object 
 \textit{ip\_addr }. The two stateful objects have a list as container since we need to track 
 all values. In the SecSip language,  the different objects are expressed by the rules depicted in 
 the Figure \ref{fig:collections}.

\begin{figure*}[htc]
\begin{small}
\begin{verbatim}
SecSip hold:from_list=list[FIELDS:sip.from]
SecSip hold:from_list.ip_addr=list[FIELDS:sip.from.addr]
\end{verbatim}
\end{small}
\caption{Declaration of stateful objects to track the \textit{From} field values and their respective source IP addresses.}
\label{fig:collections}
\end{figure*}

 When a request of type BYE is seen by SecSip, its \textit{From} field
 is checked against  the data store object $from\_list$.    If its
 value is not in the list then the message is dropped by the
 SecSip  runtime.  The   set of  rules  that  detects   this attack is
 depicted in Figure \ref{bye.attack.rules}.
\begin{figure*}[htp]
\centering
\begin{small}
\begin{verbatim}
SecSip "FIELDS:sip.method" "!^BYE$" hold:from_list=set[FIELDS:sip.from]
SecSipRule "FIELDS:sip.method" "^BYE$" && "FIELDS:sip.from" "!@in from_list" drop
\end{verbatim}
\end{small}
\caption{SecSip rules to detect the BYE attack and drop the malicious message.}
\label{bye.attack.rules}
\end{figure*}

\paragraph{INVITE request flooding attack detection}

 Another type of DoS attack, used to illustrate the SecSip language, is
 the flooding attack that targets  either  SIP phones or proxies and
 servers. The  objective of the attack is  to  exhaust resources (CPU,
 memory,  bandwidth) of   the targeted device  by  generating multiple
 calls within a short duration of time. 

 In \cite{Chen:06},  the author proposes a  method based on thresholds
 to detect this kind of attacks. He considers  that an upper bound on
 the number  of allowed  transactions per node  should  be defined and
 enforced to  defend SIP devices  from flooding attacks.  This defense
 strategy can be easily specified in and enforced by our framework as
 depicted in Figure \ref{flooding.attack.rules}.

\begin{figure*}[htp]
\centering
\begin{small}
\begin{verbatim}
SecSipaction hold:tr=set[FIELDS:sip.via.branch]
SecSip FIELDS:sip.method "^INVITE$" declare:tr.count=counter[10;60]
SecSip tr.count "@gt 15" drop
\end{verbatim}
\end{small}
\caption{SecSip rules to detect a flooding attack INVITE transactions and drop the malicious messages.}
\label{flooding.attack.rules}
\end{figure*}

 The illustrated rule set tracks INVITE attempts within each monitored
 transaction expressed by   the    \textit{TR}  object  and     its
 \textit{count}  sub-object. In the first rule, we declare the stateful object 
that will hold transactions identified by the branch parameter from the \textit{Via} field.
The second rule declares a counter and updates it every time an Invite message is seen. 
If  the screened traffic crosses the  threshold  of  more than 15  attempts in  1  minute, SecSip will drop
 subsequent transactions.

\section{Framework evaluation}
\label{sec:evaluation}

 When we   designed and built    the SecSip  runtime, we  had  several
 goals, the most important one being efficiency in terms of both safety
 and performance.

\subsection{SecSip safety}

 Since SecSip is  designed   to ensure   SIP devices  security   in an
 adversarial environment, it is imperative that it does not introduces
 new source  of  failure and vulnerabilities.  Given  the fact that we
 implemented  the environment in  C, we employed several techniques to
 make it safe. They are listed below.

\subsubsection{Buffering optimization}

 The  stateful nature of SecSip, exposes   it to state-holding attacks
 \cite{Paxon:2001}.  Such attacks  may occur when buffering data  from
 parsed SIP message.   To prevent the  environment from  such attacks,
 each stateful  object used  by SecSip  rules is hold  within  a timed
 container that when expired,  removes  the object and blocks  further
 traffic  on the object.   The lifetime of   the container is adjusted
 according the scope of the object  (message, transaction or session).
 Furthermore, each buffered  data   is normalized  to a  maximum  size
 specified by the user from SecSip  rules.  For example, the following
 SecSip rule normalizes a buffered SIP URI object to a maximum size of
 1024 bytes.  Parsers do strictly limit  the length of data at runtime
 to avoid buffer overflows.

\begin{small}
\begin{verbatim}
SecSip FIELDS:sip.uri "@normalize 1024"
\end{verbatim}
\end{small}

\subsubsection{SIP parser optimization}

In the SecSip runtime, we employed a  lazy parser to optimize the time
consumed by the parsing of a SIP message.  After  loading   the rules, the
SecSip runtime computes which objects  of a SIP message are referenced
by  the  user defined rules.   Therefore,  only those objects  will be
defined in the parse tree generated by the SIP parser module.

\subsection{SecSip performance}

 SIP dialogs and transactions are delay  sensitive.  Therefore a SIP
 firewall needs  low overhead  when inspecting packets before
 forwarding them to the target devices. 
 To assess the performance of our system, we start by looking at how
 SecSip performs with a stressing SIP traffic and what latency it adds
 while inspecting this traffic.

 The testing environment is composed of three  hosts with a core 2 CPU
 cadenced at 2.93GHZ with 2 GB of RAM. The first host plays the role of
 the SIP packet injector where we have used the SIPp tool \cite{sipp}.
 This tool  is  dedicated to sip performance   testing  and provides a
 flooding feature capable  to generate INVITE SIP messages
 at higher rates. 
 The second host, where SecSip is running, is deployed as as a bump in
 the wire device  between the attacking host  where the INVITE messages
 injector  is  deployed and the  targeted  host.  The  three hosts are
 connected through a 100 Mbits  switched Ethernet. SecSip performs all necessary
 functions including  (1) capturing  the  incoming SIP traffic on  the
 specified interface (2)  extracting SIP fields, (3) Comparing
 extracted fields with the  available rules of the current transaction
 phase, and forwarding the  SIP  messages to the outgoing  interface
 towards SIP  devices.  Our metrics  are the  delay introduced  by the
 SecSip runtime  while processing a SIP message  and its throughput in
 terms of messages/second.  To measure the delay introduced by SecSip,
 a Tcpdump instance   did  continuously run   on  the incoming-traffic
 interface to  capture incoming SIP INVITE  messages from the attacker
 and  another Tcpdump instance was operational  on the egress interface
 to  capture  the same  messages after   having  been processed by the
 SecSip runtime.  We  analysed  captured data  using a developed  Perl
 scripts to compute  the throughput and  processing delays of  SecSip.

In this work, we evaluated the performance of SecSip with two scenarios:
\begin{itemize}
\item In the first one, we only vary  the rate of INVITE messages
 from 10 to  1000 messages/second  by a  step   of 10.  Each value  is
 maintained by the SIPp tool one minute and then  it is incremented by
 10. SecSip is deployed without any rule, only messages parsing overhead
is measured. 
\item In the second one, we maintained the INVITE message rate at a level of 60 messages
per second. We then vary for each test the number of rules from 1 to 650 by a step of 10.
Herein, our goad is to measure the overhead of the rules processing module.
\end{itemize}

\begin{figure}[htc]
 \centering
\scalebox{0.4}{\includegraphics{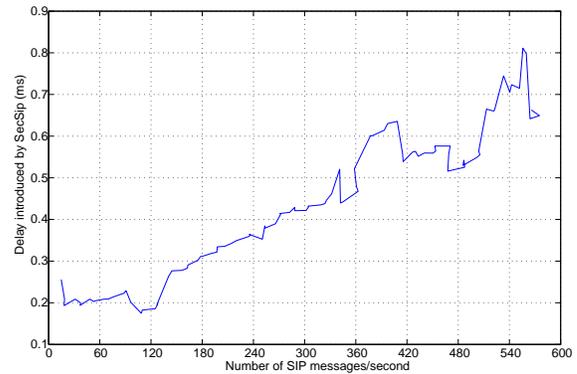}}
\caption{Delays introduced by SecSip deployed as a bump in the wire device. No rules are loaded by SecSip}
\label{fig:delays}
\end{figure}

\begin{figure}[htc]
 \centering
\scalebox{0.4}{\includegraphics{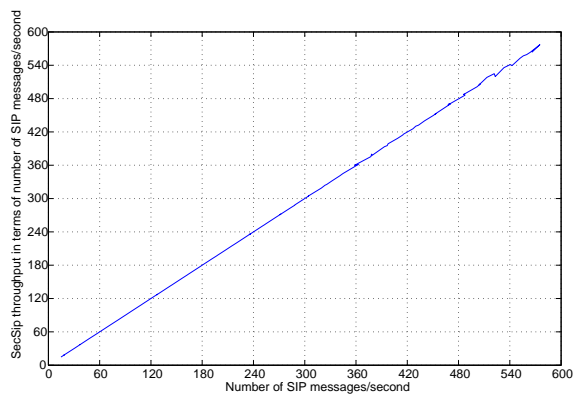}}
\caption{Throughput of SecSip in terms of number of SIP messages per second. No rules are loaded by SecSip.}
\label{fig:throughput}
\end{figure}

The results of the first scenario are depicted in Figures \ref{fig:delays} and
\ref{fig:throughput}. In this scenario, we measure the effect of a
 stressing load towards SecSip.
 In figure \ref{fig:delays}, we show  the mean delay introduced by the
 SecSip runtime  while processing a  varying  number of injected INVITE
 messages towards the SIP device. We observe that the delay  stays
 below 1 ms, even with an injection rate close to 500 messages/second.

 The SecSip throughput is  depicted in Figure  \ref{fig:throughput}
 with respect to an injection rate from the source host. We observe  that the throughput 
 stays close to the injection rate. 
 
\begin{figure}[htc]
 \centering
\scalebox{0.4}{\includegraphics{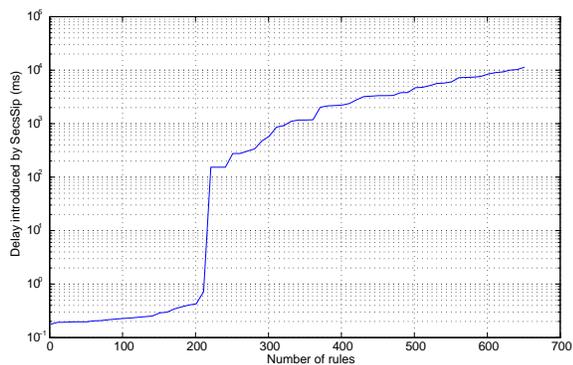}}
\caption{Delays introduced by SecSip deployed as a bump in the wire device. 
We varied the number of rules loaded by SecSip. The y-axis is in log scale.}
\label{fig:delays.rules}
\end{figure}

\begin{figure}[htc]
 \centering
\scalebox{0.4}{\includegraphics{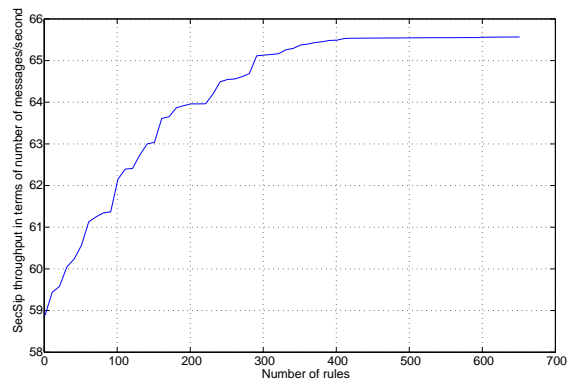}}
\caption{Throughput of SecSip in terms of number of SIP messages per second. 
We varied the number of rules loaded by SecSip.}
\label{fig:throughput.rules}
\end{figure}
The results of the second scenario, where we only varied the number of rules,
are depicted in Figures \ref{fig:delays.rules} and \ref{fig:throughput.rules}.
We observe from the Figure \ref{fig:throughput.rules} that SecSip maintains its throughput close
to the injection rate of 60 SIP messages/second. 
However, as depicted in Figure \ref{fig:delays.rules}, the delays introduced by SecSip 
become more important when we increase the number of loaded rules. 

It seems that the rules processing module is the largest component contributing to the SecSip overhead.
Therefore, we need  to better
optimize this component, to obtain a better  performance of the firewall.

\section{Conclusions and Future work}
\label{sec:conclusion}

 With  the increasing importance of SIP-based  systems  in the Internet,
 the availability of defense solutions able to protect all these systems
 against malicious exploitation  of vulnerabilities is  essential.  In
 this paper, we  have shown that SecSip is  one solution able to deal
 with    known,  and to  some    extend  unknown,  vulnerabilities  by
 efficiently building  a per device tuned  protection scheme.  Our key
 contribution  include the  design  of  a  runtime  and  a  rule-based
 language  to     protect   SIP-based    networks  from     discovered
 vulnerabilities.  The specification language  is easy to use for authoring
 SIP vulnerabilities based on the protocol  states prior any potential
 exploitation, along  with message parsing for  exploit detection.  To
 achieve this, the SecSip language  allows the use of stateful objects
 that track protocol      states.    The evaluation  of      a  SecSip
 implementation, indicates that the  introduced delays  are acceptable
 for an   on-line analysis  engine.    The   SecSIP  implementation   is
 distributed in Open Source (GPL 2 license) and  can be downloaded from
 the INRIA Gforge \footnote{http://secsip.gforge.inria.fr}.

 We  are currently working   on     the improvement of  the     SecSip
 implementation and its language  to stabilize the release and provide
 improved  performances.  Interfaces with  several management services
 are also under development (e.g. Syslog, SNMP). 
 The idea behind it is to allow secsip to interact directly with SIP devices 
  management interfaces  to collect some useful data. For example, SecSip may need to know if the SIP device is up 
 or down.  Finally, we also plan 
 to couple SecSip with attack and vulnerability tools \cite{kiph} to automatically
 generate  defense rules.

\bibliographystyle{IEEEtran}
\bibliography{IEEEabrv,im09}

\begin{thebibliography}{10}
\providecommand{\url}[1]{#1}
\csname url@samestyle\endcsname
\providecommand{\newblock}{\relax}
\providecommand{\bibinfo}[2]{#2}
\providecommand{\BIBentrySTDinterwordspacing}{\spaceskip=0pt\relax}
\providecommand{\BIBentryALTinterwordstretchfactor}{4}
\providecommand{\BIBentryALTinterwordspacing}{\spaceskip=\fontdimen2\font plus
\BIBentryALTinterwordstretchfactor\fontdimen3\font minus
  \fontdimen4\font\relax}
\providecommand{\BIBforeignlanguage}[2]{{%
\expandafter\ifx\csname l@#1\endcsname\relax
\typeout{** WARNING: IEEEtran.bst: No hyphenation pattern has been}%
\typeout{** loaded for the language `#1'. Using the pattern for}%
\typeout{** the default language instead.}%
\else
\language=\csname l@#1\endcsname
\fi
#2}}
\providecommand{\BIBdecl}{\relax}
\BIBdecl

\bibitem{rfc:sip}
\BIBentryALTinterwordspacing
J.~Rosenberg, H.~Schulzrinne, G.~Camarillo, A.~Johnston, J.~Peterson,
  R.~Sparks, M.~Handley, and E.~Schooler, ``{SIP: Session Initiation
  Protocol},'' RFC 3261 (Proposed Standard), Jun. 2002, updated by RFCs 3265,
  3853, 4320, 4916. [Online]. Available:
  \url{http://www.ietf.org/rfc/rfc3261.txt}
\BIBentrySTDinterwordspacing

\bibitem{voipsa}
VoIPSA.org, ``{VOIPSEC} mailing list on {VoIP} security issues,''
  {http://voipsa.org/mailman/listinfo/voipsec\_voipsa.org}, January 2009.

\bibitem{zhang:07}
R.~Zhang, X.~Wang, X.~Yang, and X.~Jiang, ``{B}illing attacks on {SIP}-based
  {V}o{IP} systems,'' in \emph{WOOT '07: Proceedings of the first {USENIX}
  workshop on Offensive Technologies}.\hskip 1em plus 0.5em minus 0.4em\relax
  Berkeley, CA, USA: USENIX Association, 2007, pp. 1--8.

\bibitem{state:07}
H.~Abdelnur, R.~State, I.~Chrisment, and C.~Popi, ``{A}ssessing the security of
  {V}o{IP} {S}ervices,'' in \emph{Integrated Network Management, IM 2007. 10th
  IFIP/IEEE International Symposium on Integrated Network Management, Munich,
  Germany, 21-25}.\hskip 1em plus 0.5em minus 0.4em\relax IEEE, May 2007, pp.
  373--382.

\bibitem{Chen:06}
E.~Chen, ``Detecting dos attacks on sip systems,'' \emph{VoIP Management and
  Security, 2006. 1st IEEE Workshop on}, pp. 53--58, April 2006.

\bibitem{Sip:threats}
D.~Geneiatakis, T.~Dagiuklas, G.~Kambourakis, C.~Lambrinoudakis, S.~Gritzalis,
  K.~Ehlert, and D.~Sisalem, ``Survey of security vulnerabilities in session
  initiation protocol,'' \emph{Communications Surveys \& Tutorials, IEEE},
  vol.~8, no.~3, pp. 68--81, Qtr. 2006.

\bibitem{Habib:03}
A.~Habib, M.~M. Hefeeda, and B.~K. Bhargava, ``Detecting service violations and
  dos attacks,'' in \emph{In Proceedings of 2003 Internet Society Symposium on
  Network and Distributed System Security (NDSS’03}, 2003, pp. 177--189.

\bibitem{modsecurity}
I.~Rustic, \emph{ModSecurity Reference Manual v2.5.5}, Breach Security, Juin
  2008.

\bibitem{snort}
Snort, \emph{The Open Source Network Intrusion Detection System},
  http://www.snort.org, Juin 2008.

\bibitem{bro}
V.~Paxson, ``Bro: a system for detecting network intruders in real-time,''
  \emph{Comput. Netw.}, vol.~31, no. 23-24, pp. 2435--2463, 1999.

\bibitem{fielder:07}
J.~Fielder, T.~Kupta, S.~Ehlert, T.~Magedanz, and D.~Sisalem, ``{VoIP
  Defender}: Highly scalable sip-based security architecture,'' in
  \emph{International Conference on Principles, Systems and Applications of IP
  Telecommunications (IPTComm)}, ACM, Ed., New York, USA, 19-20 July 2007, pp.
  11--17, iSBN: 978-1-60558-006-7.

\bibitem{Wu:04}
Y.-S. Wu, S.~Bagchi, S.~Garg, N.~Singh, and T.~Tsai, ``Scidive: A stateful and
  cross protocol intrusion detection architecture for voice-over-ip
  environments,'' in \emph{DSN '04: Proceedings of the 2004 International
  Conference on Dependable Systems and Networks}.\hskip 1em plus 0.5em minus
  0.4em\relax Washington, DC, USA: IEEE Computer Society, 2004, p. 433.

\bibitem{sengar:06}
H.~Sengar, D.~Wijesekera, H.~Wang, and S.~Jajodia, ``Voip intrusion detection
  through interacting protocol state machines,'' \emph{Dependable Systems and
  Networks, 2006. DSN 2006. International Conference on}, pp. 393--402, 2006.

\bibitem{kiph}
H.~Abdelnur, O.~Festor, and R.~State, ``Kif: A stateful sip fuzzer,'' in
  \emph{1st International Conference on Principles, Systems and Applications of
  IP Telecommunications (IPTComm)}, ACM, Ed., July 2007.

\bibitem{Paxon:2001}
M.~Handley, V.~Paxson, and C.~Kreibich, ``Network intrusion detection: evasion,
  traffic normalization, and end-to-end protocol semantics,'' in \emph{SSYM'01:
  Proceedings of the 10th conference on USENIX Security Symposium}.\hskip 1em
  plus 0.5em minus 0.4em\relax Berkeley, CA, USA: USENIX Association, 2001, pp.
  9--9.

\bibitem{sipp}
R.~Gayraud and O.~Jacques, \emph{SIPp Reference Manual}, Juin 2008,
  http://sipp.sourceforge.net/.

\end{thebibliography}
\end{document}